\documentstyle[12pt]{article}
\def\bfl{\begin{flushleft}}
\def\efl{\end{flushleft}}
\def\bfr{\begin{flushright}}
\def\efr{\end{flushright}}
\def\bc{\begin{center}}
\def\ec{\end{center}}
\def\be{\begin{equation}}
\def\ee{\end{equation}}
\def\ba{\begin{eqnarray}}
\def\ea{\end{eqnarray}}

\def\text#1{\mbox{#1}}
\def\drm{\text{d}}
\def\Sign#1{\, \text{sign}\left(#1\right) }
\def\Eq{eq.~}
\def\Ref{ref.~}

\begin{document}

~\\
\bfr
LANL e-print gr-qc/9802041\\
Int. J. Mod. Phys. D 8 (1999) 549-555\\
\efr
~~\\

\bc
{\Large \bf
Barotropic thin shells with linear EOS as models of 
stars and circumstellar shells in general relativity
}

~~\\
Konstantin G. Zloshchastiev\\
~~\\
Box 2837, Dnepropetrovsk 320128, Ukraine; 
e-mail: zlosh@email.com
\ec

~\\

\abstract{
The spherically symmetric thin shells of the barotropic fluids with the
linear equation of state  
are considered within the frameworks of general relativity. 
We study several aspects of the shells as 
completely relativistic models of 
stars, first of all the neutron stars and white dwarfs, and circumstellar
shells.
The exact equations of motion of the shells are obtained.
Also we calculate the parameters of the equilibrium configurations,
including the radii of static shells.
Finally, we study the stability of the equilibrium shells against radial 
perturbations.
}

~\\

PACS number(s):  04.40.Nr, 11.27.+d, 12.38.Mh, 97.10.Fy

Keyword(s): general relativity, thin shell, star, circumstellar shell\\

~~\\

\newpage

In present paper we consider some infinitely thin (singular) shells 
as models of the physical entities, which thickness is negligible in 
comparison with a  circumference radius
(e.g., the circumstellar shells or surfaces of phase transitions).
Besides, one may study the shells as the simplest (nevertheless useful and 
instructive) models of ``thick'' objects, e.g., stars.
Geometrically the shell is described by a three-dimensional closed 
hypersurface, embedded in
the four-dimensional spacetime and dividing it into the two domains, 
the external ($\Sigma^+$) and internal ($\Sigma^-$) spacetime.
Since the classical works \cite{dau,isr} the theory 
of surface layers has been widely considered in the literature 
(see \Ref \cite{mtw} for details) including modern applications and 
developments in fundamental, quantum and phenomenological directions,
see \Ref \cite{zlo} and references therein.
We only point out some questions of principle now.
One of them is the essential difference between the description of the 
boundary surface (for instance, surface of star) and shell.
The boundary surface is the discontinuity of the first kind 
(mass density has a finite jump across the surface) 
and is described by the Lichnerowicz-Darmois junction conditions
(the first and second quadratic forms are continuous on the surface).
The singular thin shell is the discontinuity of the second kind
(the density has the delta-like singularity on the shell) and is 
described by the Lichnerowicz-Darmois-Israel junction conditions:
the first quadratic form (the metric) is continuous, the second one
(the extrinsic curvature) has a finite jump.

This circumstance says we must take care at physical interpretation
(and application) of the shells.
For instance, the surface of a star should be described by a boundary 
surface not a singular thin shell.
Latter can be used as a model: (a) of the star as a whole system \cite{nun}, 
we replace a thick layer of matter (which has an infinite number of 
degrees of freedom) by the singular shell which has a finite number of 
collective degrees of freedom (spherically symmetric thin shells 
have only radial degree of freedom), 
(b) of the stellar internal structure, thereby a star can 
be considered as a set of an infinite number of embedded shells,
(c) of the circumstellar shells as a set of a finite number of shells.
The aim of this paper is to study all appropriate cases for the simplest 
class of equations of state (EOS).

So, one considers a thin shell with the surface stress-energy 
tensor of a perfect fluid in the general case 
\be
S_{ab}=\sigma u_a u_b + p (u_a u_b +~ ^{(3)}\!g_{ab}),
                                                               \label{eq1}
\ee
where $\sigma$ and $p$ are the surface energy density and 
pressure respectively, $u^a$ is the timelike unit tangent vector, 
$^{(3)}\!g_{ab}$ is the 3-metric of a shell's surface.
We suppose metrics of the spacetimes outside $\Sigma^+$ and inside 
$\Sigma^-$ of a spherically symmetric shell to be of the 
Schwarzschild form, i.e.,
\be
\drm s_\pm^2 =
-\left[
       1-\frac{2M_\pm}{r}
\right] \drm t^2_\pm + 
\left[ 1 -\frac{2M_\pm}{r}
\right]^{-1} \drm r^2 + r^2 \drm \Omega^2,                     \label{eq2}
\ee
where $\drm \Omega^2$ is the metric of the unit 2-sphere.
It corresponds to the neutral shell surrounding a neutral body of 
mass $M_-$, and $M_+$ appears to be the external observable total mass 
(energy) of the whole system.
We assume both masses to be non-negative (though a theory of quantum 
gravity could give rise to spacetimes of negative mass, 
see \Ref \cite{zlo,man}).
It is possible to show that if one uses the  proper time  
$\tau$ then the 3-metric of a shell is
\be
^{(3)}\!\drm s^2 = - \drm \tau^2 + R^2 \drm \Omega^2,        \label{eq3}
\ee
where $R=R(\tau)$ is a proper radius of a shell.
Also the energy conservation law for matter on a shell 
(which is an integrability condition for the Einstein equations on a shell)
can be written as
\be
\drm \left( \sigma ~^{(3)}\!g \right) +
p~ \drm \left( ~^{(3)}\!g \right)  =0,                       \label{eq4}
\ee
where  $^{(3)}\!g=\sqrt{-\det{(^{(3)}\!g_{ab})}} = R^2 \sin{\theta}$.
In this equation, the first term corresponds to a change in
the shell's
internal energy, the second term corresponds to the work done
by the shell's internal forces.

Imposing junction conditions across a shell, the Einstein equations
yield the equation of shell's motion in the form
\be
\epsilon_+ \sqrt{1+\dot R^2 - \frac{2 M_+}{R}} - 
\epsilon_- \sqrt{1+\dot R^2 - \frac{2 M_-}{R}} = - \frac{m}{R}, \label{eq5}
\ee
where
\be
m = 4 \pi \sigma R^2                                      \label{eq6}
\ee
is interpreted as an effective rest mass, $\dot R=\drm R/\drm\tau$ is the 
proper velocity of a shell, 
$\epsilon_\pm = \Sign{\sqrt{1+\dot R^2 - 2 M_\pm/R}}$.
Equations (\ref{eq4}) - (\ref{eq6}) together 
with an equation of state and choice of the 
signs $\epsilon_\pm$, completely determine the motion of the fluid 
Schwarzschild shell in general relativity.
Therefore, we must resolve the next two problems, viz., the choice of 
$\epsilon_\pm$ in \Eq (\ref{eq5}), and the choice of EOS.

(i) {\it The choice of signs $\epsilon_\pm$}.
It is well-known, that $\epsilon = +1$ if $R$ increases in 
the outward normal direction to the shell ( e.g., it 
takes place in a flat spacetime),
and $\epsilon = -1$ if $R$ decreases (semiclosed world).
Thus, only under the additional condition $\epsilon_+ = 
\epsilon_-=1$ we have an ordinary shell \cite{bkt,gk}.
In the paper we model ordinary stars not the wormhole-stars, therefore,
in (\ref{eq5}) we suppose $\epsilon_+ = \epsilon_-=1$.

(ii) {\it The choice of the equation of state}.
We confine our discussion to models calculated with ultra-high density
barotropic state equations of the linear form
\be
^{(3)}\!p = (\gamma - 1) ~^{(3)}\!\varepsilon.              \label{eq7}
\ee
For physical purposes the cores with $1<\gamma \leq 2$, e.g., 
$\gamma=\, 2,\, 4/3,\, 6/5,\, 14/13$ were considered \cite{mz} as
models of neutron stars.
They showed a maximum mass of approximately $0.7\,M_\odot$
corresponding to a radius around four Schwarzschild radii. 
For two-dimensional matter \Eq (\ref{eq7}) implies
\be
p = \eta \sigma,                                           \label{eq8}
\ee
where the reduction of dimensionality of the stress-energy tensor
should be taken into account, i.e., $2 \eta = 3 (\gamma - 1)$.
This equation includes the most studied cases: the dust shell $p=0$
\cite{isr,hb},
radiation fluid shell $\sigma - 2 p=0$ \cite{vis}, and bubble 
$\sigma + p =0$ \cite{bkt,cl}.
If $\eta >0$ it can be interpreted as a square component of the vector of a 
speed of sound in the shell.
Then for a spatially two-dimensional homogeneous fluid the
square speed of sound is $2 \eta$.
From the physical viewpoint some $\eta$ appear to be inadmissible.
For instance, if the fluid on a shell is required to satisfy the dominant 
energy condition, $\sigma \geq |p|$, one obtains the constraint
\be
|\eta| \leq 1.                                             \label{eq9}
\ee
If a fluid is required to satisfy the causality condition,
we get the constraint
\be
\eta \leq 1/2,                                              \label{eq10}
\ee 
where one took into account spatial two-dimensionality of the fluid.
Nevertheless, the aim of this paper is to study the general case of 
arbitrary $\eta$ \cite{zlo002}.
So, solving the differential equation (\ref{eq4}) with respect to $\sigma$,
we obtain
\be
\sigma  = \frac{C}{4\pi} R^{-2 (\eta + 1)},                \label{eq11}
\ee
where $C$ is the integration constant determined by the specific shell
matter.
The value of $C$ is closely related to the value of surface mass density
(or pressure) at fixed $R$.
We consider ordinary shells not wormholes, therefore, we require 
$\sigma > 0$, hence $C>0$.
Also it should be noted that from (\ref{eq6}) at positive 
densities $\sigma$ it follows 
that $M_+ > M_-$ for any $R$ and $\dot R$.
Otherwise, matching of the spacetimes (\ref{eq2}) is impossible.
Thus, we have all necessary equations (\ref{eq5}), (\ref{eq6}) and 
(\ref{eq11}) to study the shells as models of stars.

First of all we study the static shells.
Taking into account (\ref{eq8}), the equilibrium 
conditions $\dot R=\ddot R=0$ \cite{fhk} for our case read
\be
k_- - k_+ = 4\pi\sigma_0 r_0,                         \label{eq12}
\ee
\be
\frac{M_+}{k_+} - \frac{M_-}{k_-} = 
4\pi (2 \eta +1) \sigma_0 r_0^2,                           \label{eq13}
\ee
where $k_\pm = \sqrt{1-2 M_\pm/r_0}$ ($k_+ < k_-$), 
$r_0$ and $\sigma_0$ are the shell
radius $R$ and density at equilibrium respectively, 
$\sigma_0$ is given by (\ref{eq11}) at $R=r_0$.
Taking into account (\ref{eq12}), the equation (\ref{eq13}) can be
rewritten in the nice form
\be
\frac{1}{k_+ k_-} = 4 \eta +1.                               \label{eq14}
\ee
We can use the system (\ref{eq12}), (\ref{eq14}) instead (\ref{eq12}), 
(\ref{eq13}).
The equation (\ref{eq14}) is very helpful because it already gives us
both the radius $r_0$ and constraint for $\eta$ independently
of the arbitrary constant $C$.

{\it Proposition.}
Let we have the neutral ordinary shell of the matter with the 
EOS (\ref{eq8}), $k_\pm$ are real and positive, $M_+ \not= M_-$.
Then the shell can have static states only at
\[
\eta > 0.
\]

{\it Proof:} 
From \Eq (\ref{eq14}) it follows that 
$4 \eta + 1 > 0$,
and
\[
1- \frac{ (4 \eta + 1)^{-2} }{ 1- 2 M_-/r_0} = \frac{2 M_+}{r_0} >0,
\]
or 
\[
(4 \eta + 1)^{-2} < 1.
\]
These conditions give the desired inequality $\eta > 0$, Q.E.D.

Hence it follows that neither the dust shells ($\eta = 0$), nor bubbles 
($\eta = -1$) can have static states with mass $M_+$.
In other words, they should not be considered as {\it static} sources of the 
Schwarzschild metric (\ref{eq2}).
By means of this proposition we can prove the next theorem.

{\it Theorem.}
Let the conditions of the previous proposition are valid.
Then the radius of static shells is given by the expression
\be
r_0 = 
\frac{1}{\varrho}
\left[
       M_+ + M_- + \sqrt{\Delta M^2 + 4 M_+ M_- (1-\varrho)}
\right],                                                       \label{eq15}
\ee
where $\varrho = 1 - (4 \eta + 1)^{-2}$, $\Delta M = M_+ - M_-$.

{\it Proof:} 
Perform the conformal transformation $\xi = r_0/2 M_+$, $\alpha = M_-/M_+$.
The case $\alpha = 0$ corresponds to the hollow shell ($M_- = 0$),
but $\alpha = 1$ describes the limit case 
$M_\pm \rightarrow \infty$, $\Delta M \rightarrow 0$ 
(infinitely light shells), 
not the case $M_+ \equiv M_-$ 
which for ordinary shells is forbidden as trivial (\ref{eq5}).
Therefore $0 \leq \alpha \leq 1$.
Then, squaring \Eq (\ref{eq14}), we obtain the square equation with
respect to $\xi$
\[
\varrho \xi^2 - (1+ \alpha) \xi + \alpha = 0,
\]
which has the two roots
\be
2 \varrho\, \xi_\pm = 
1 + \alpha \pm \sqrt{(1 + \alpha)^2 - 4 \alpha \varrho},
\label{eq16}
\ee
one of them is superfluous, as can be directly verified.
The preceding proposition yields 
\[
0<\varrho < 1,~
\sqrt{(1 + \alpha)^2 - 4 \alpha \varrho} < 1 + \alpha.
\]
Further, taking into account \Eq (\ref{eq16}), the condition $\xi > 1$
can be written as
\be
\pm \sqrt{(1 + \alpha)^2 - 4 \alpha \varrho} > 2\varrho - 1 - \alpha.
\label{eq17}
\ee
If we define the auxiliary function
\[
z_\pm (\varrho) = 
1 + \alpha -2\varrho \pm \sqrt{(1 + \alpha)^2 - 4 \alpha \varrho},
\]
then we must require $z_\pm (\varrho) > 0$.
It can easily be seen that 
\[
z_+ (\varrho \not = 1) >0,~ z_+ (1) = 0,~
z_- (\varrho \not = 0) <0,~ z_- (0) = 0,
\]
therefore, only the root $\xi_+$ is admissible, Q.E.D.

The radius $r_0$ given by \Eq (\ref{eq15}) is
illustrated in Fig.~\ref{fig1} for the most physically
admissible range, 
$0\, \text{(dust)} <\eta \leq 1/2 \, \text{(ultrarelativistic fluid)}$.
For clarity only the two limit cases, $\alpha=0$ and $\alpha=1$, 
are represented, 
the curves for other $0<\alpha<1$ lie between them.
The parameter $1-\alpha = \Delta M/ M_+$ determines the proper {\it total}
mass of the shells.
As it was mentioned above, we can model both the whole star and stellar
internal structure.
In this connection the model with $\alpha=0$ (the hollow shell) appears
to be a model
of the star as a whole system, whereas the shells at $\alpha \not =0$ can
be models 
both inhomogeneous stellar structure (the layers with 
smaller $\eta$ lie above those with greater $\eta$) 
and the circumstellar shells.

Another important aspect we must consider is the mechanical stability
of the shells.
Here we apply the method developed in \Ref \cite{fhk}.
Redefining the constant $C$ in (\ref{eq11}), we suppose
\be
\frac{\sigma}{\sigma_0} = \frac{1}{4\pi}
z^{-2 (\eta +1)},                                  \label{eq18}
\ee
where $z=R/r_0$, the constants $\sigma_0$ and $r_0$ should be calibrated
by means of the system (\ref{eq12}), (\ref{eq14}).
It is well-known that the shell's equation of motion can be rewritten 
in a more convenient form.
Squaring (\ref{eq5}) twice, one obtains
\be
\dot R^2 + V (R) =0,                                  \label{eq19}
\ee
where the potential $V(R)$ is
\[
V(R) = 1 - \frac{2 M_+}{R} -
\left[
      \frac{a}{R}
      \left(
            \frac{R}{r_0}
      \right)^{2 \eta +1}
      -
      \frac{\Delta k}{2}
      \left(
            \frac{r_0}{R}
      \right)^{2 \eta +1}
\right]^2,                                       
\]
where $a = \Delta M/ \Delta k$, $\Delta k = k_+ - k_-$.
Calibrating these expressions, we obtain
\be
r_0^2 \dot z^2 + V (z) =0,                                   \label{eq20}
\ee
where
\[
V(z) = 1 + \frac{k_+ -1}{z} -
\left[
  \frac{
        \varrho (k_+^2 + 1) z^{4\eta +1}
        +
        \varrho k_+^2 - 1
       }
       {
         2 \varrho k_+ z^{2\eta +1}
       }
\right]^2.                                                 
\]
The potential satisfies $V(z=1) = \drm V / \drm z|_{z=1} = 0$, as can 
be directly verified.
Therefore, the stability of the equilibrium is equivalent to $V(z)$
having a minimum at $z=1$.
By differentiation one obtains
\be
\left.
\frac{\drm^2 V}{\drm z^2}
\right|_{z=1} = - 
               \frac{k_+^4 (4\eta+1)^2 - 2 k_+^2 +1}
                    {2 k_+^2}.                                 \label{eq21}
\ee
From \Eq (\ref{eq21}) and from the proposition proved above it can readily
be seen that $\drm^2 V/\drm z^2|_{z=1} < 0$.
Therefore, the equilibrium of the neutral shells with the linear EOS 
(unlike those with the linearized EOS studied by 
Brady, Louko and Poisson \cite{fhk}) 
is unstable against radial perturbations.
In application to models of neutron stars it means that the 
EOS (\ref{eq8}) can not purely describe them.
Apparently only the average layers of the star core can be ``made'' from
the matter (\ref{eq8}).
The central part of the core (the so called ``core of the core'') and 
core's crust must be consisting of other kind substance to prevent
from collapse into black holes and explosion respectively.
For instance, Witten \cite{wit} argued that the strange matter can be 
formed at the quark-hadron phase transition in central part of neutron 
stars. 
As for the crust that it can  be even crystallic \cite{zn}.

Thus, in the present paper some aspects of dynamics of the spherically
symmetric shells with the linear EOS were considered.
Finally it should be noted that the studied shells may also describe, 
besides static (or having adiabatically slow radial motion) 
circumstellar shells and stars, the 
transient astrophysical phenomena, e.g., the blast waves at supernova 
explosion.
Thereby the obtained models of these phenomena will be completely
relativistic that is essential at high velocities and powerful fields
when the spacetime can not be supposed flat.

\def\APP{Acta Phys. Polon.}
\def\CMPh{Commun. Math. Phys.}
\def\CJP{Czech. J. Phys.}
\def\CQG {Class. Quantum Grav.}
\def\FP{Fortschr. Phys.}
\def\GRG {Gen. Relativ. Gravit.}
\def\IJMP {Int. J. Mod. Phys.}
\def\JPh{J. Phys.}
\def\LMPh {Lett. Math. Phys.}
\def\MPL {Mod. Phys. Lett.}
\def\NPh  {Nucl. Phys.}
\def\PhE  {Phys.Essays}
\def\PhL  {Phys. Lett.}
\def\PhR  {Phys. Rev.}
\def\PhRL {Phys. Rev. Lett.}
\def\PhRp {Phys. Rep.}
\def\NCim {Nuovo Cimento}
\def\NuPB {Nucl. Phys.}
\def\prp {report}
\def\Prp {Report}

\def\jn#1#2#3#4#5{{#1}{#2} {#3} {(#5)} {#4}}   

\def\boo#1#2#3#4#5{ #1 ({#2}, {#3}, {#4}){#5}}  

\def\prpr#1#2#3#4#5{{``#1,''} {#2}{#3}{#4}, {#5} (unpublished)}

\def\And{~and~}

~~\\
~~\\
~~\\
~~\\

\begin{figure}
\caption{Plot of the static shell radius (in units of the
Schwarzschild radius) vs $\eta$. 
The dashed curve describes hollow shells ($\alpha = 0$), 
the solid curve does the asympotical case $\alpha = 1$ (infinitely light 
shells). 
The shaded region corresponds to black holes.}
\label{fig1}
\end{figure}


\newpage

\begin{thebibliography}{99}

\bibitem{dau}
G. Dautcourt, \jn{Math. Nachr.}{}{27}{277}{1964}.

\bibitem{isr}
W. Israel, \jn{\NCim}{}{44B}{1}{1966}.

\bibitem{mtw}
C. W. Misner, K. S. Thorne\And J. A. Wheeler, 
\boo{Gravitation}{Freeman}{San Francisco}{1973}{}.

\bibitem{zlo}
K. G. Zloshchastiev,
\jn{\GRG}{}{31}{571}{1999}; 
\jn{\IJMP}{ D}{8}{165}{1999}; 
\jn{{\it ibid.}}{}{8}{363}{1999}; 
\jn{\APP}{ B}{30}{897}{1999};
\jn{\CQG}{}{16}{1737}{1999}.


\bibitem{nun}
D. N{\'u}{\~n}ez, \jn{Astrophys. J.}{}{482}{963}{1997}.

\bibitem{man}
R. Mann, \jn{\CQG}{}{14}{2927}{1997}.

\bibitem{bkt}
P. Laguna-Castillo and R. A. Matzner, \jn{Phys. Rev.}{ D}{34}{2913}{1986};
V. A. Berezin, V. A. Kuzmin\And I. I. Tkachev, 
\jn{\PhR}{ D}{36}{2919}{1987}.

\bibitem{gk}
V. A. Berezin, V. A. Kuzmin\And I. I. Tkachev, 
\jn{\PhL}{ B}{120}{91}{1983}; 
V. A. Berezin, N. G. Kozimirov, V. A. Kuzmin\And I. I. Tkachev, 
\jn{\PhL}{ B}{212}{415}{1988}; 
D. S. Goldwirth and J. Katz, \jn{\CQG}{}{12}{769}{1995}.

\bibitem{mz}
C. W. Misner and H. S. Zapolsky, \jn{Phys. Rev. Lett.}{}{12}{635}{1964}.

\bibitem{hb}
P. H\'aj\'\i\v{c}ek and J. Bi\v{c}\'{a}k, 
\jn{\PhR}{ D}{56}{4706}{1997};
J. L. Friedman, J. Louko\And S. N. Winters-Hilt, 
\jn{{\it ibid.}}{}{56}{7674}{1997};
K. G. Zloshchastiev, 
\jn{{\it ibid.}}{}{57}{4812}{1998}.

\bibitem{vis}
M. Visser,  
\jn{Nucl. Phys.}{ B}{328}{203}{1989}.

\bibitem{cl}
S. Coleman and F. De Luccia, \jn{Phys. Rev.}{ D}{21}{3305}{1980}.

\bibitem{zlo002}
K. G. Zloshchastiev,
\jn{\MPL}{ A}{13}{1419}{1998}.


\bibitem{fhk}
J. Frauendiener, C. Hoenselaers\And W. Konrad, 
\jn{\CQG}{}{7}{585}{1990};
P. R. Brady, J. Louko\And E. Poisson, 
\jn{\PhR}{ D}{44}{1891}{1991}. 

\bibitem{wit}
E. Witten, \jn{\PhR}{ D}{30}{272}{1984}. 

\bibitem{zn}
Ya. B. Zel'dovich and I. N. Novikov, 
\boo{Theory of gravity and evolution of stars}{Nauka}{Moskow}{1971}{}.

\end{thebibliography}
\end{document}